\newcommand{\mevcc}{\!\mathrm{MeV}/c^2}

\newcommand{\mev}{\!\mathrm{MeV}}
\newcommand{\gevcc}{\!\mathrm{GeV}/c^2}

\newcommand{\pb}{\!\mathrm{pb}}
\newcommand{\ipb}{\!\mathrm{pb^{-1}}}

\newcommand{\sss}{\!\mathrm{\sigma}}

\documentclass{ws-ijmpcs}

\begin{document}

\markboth{Mitchell}
{RECENT RESULTS FROM CLEO-c}

%
\catchline{}{}{}{}{}
%

\title{RECENT RESULTS ON CHARMONIUM DECAYS AT CLEO-c}

\author{R.~E.~Mitchell}

\address{Department of Physics, Indiana University \\
727 E. Third St., Bloomington, Indiana 47405, USA\\
remitche@indiana.edu}

\maketitle

\begin{history}
\received{15 February 2011}
\revised{15 February 2011}
\end{history}

\begin{abstract}
The CLEO-c Experiment has made several recent contributions to the study of charmonium decays.  This review briefly outlines the CLEO-c analyses of charmonium that were completed or made public during~2010.  Special attention is given to the discovery of the process $e^+e^-\to\pi^+\pi^-h_c$ at a center of mass energy of $4170~\mev$.
\end{abstract}

\ccode{PACS numbers: 14.40.Pq}

\section{Introduction}	

Despite the conclusion of data-taking in~2008, the CLEO-c Experiment -- which utilizes symmetric $e^+e^-$ collisions in the charmonium region provided by the Cornell Electron Storage Ring (CESR) -- continues to have an active program in the analysis of charmonium data.  Recent CLEO-c studies include: (1)~measurements of the branching fractions of $\chi_{cJ}\to p\overline{p}\pi^0$, $p\overline{p}\eta$, and $p\overline{p}\omega$~\cite{iupp};   (2)~studies of $\psi(2S)$ decays to $\gamma p\overline{p}$, $p\overline{p}\pi^0$, and $p\overline{p}\eta$ and a confirmation of the BESII observation of the $X(1870)$, a $p\overline{p}$ threshold enhancement in $J/\psi\to\gamma p\overline{p}$~\cite{nupp};  (3)~a search for the radiative transition $\psi(2S)\to\gamma\eta_c(2S)$~\cite{etac};  (4)~measurements of the higher order multipoles in the radiative transitions $\psi(2S)\to\gamma\chi_{c1,2}$ and $\chi_{c1,2}\to\gamma J/\psi$~\cite{multipoles}; and (5)~the discovery of the process $e^+e^-\to\pi^+\pi^-h_c$ at a center of mass energy of $4170~\mev$ (preliminary).  In the following, each study is briefly reviewed.

\section{Branching Fractions for $\chi_{cJ}\to p\overline{p}\pi^0$, $p\overline{p}\eta$, and $p\overline{p}\omega$}

\begin{figure}[pb]
\centerline{\psfig{file=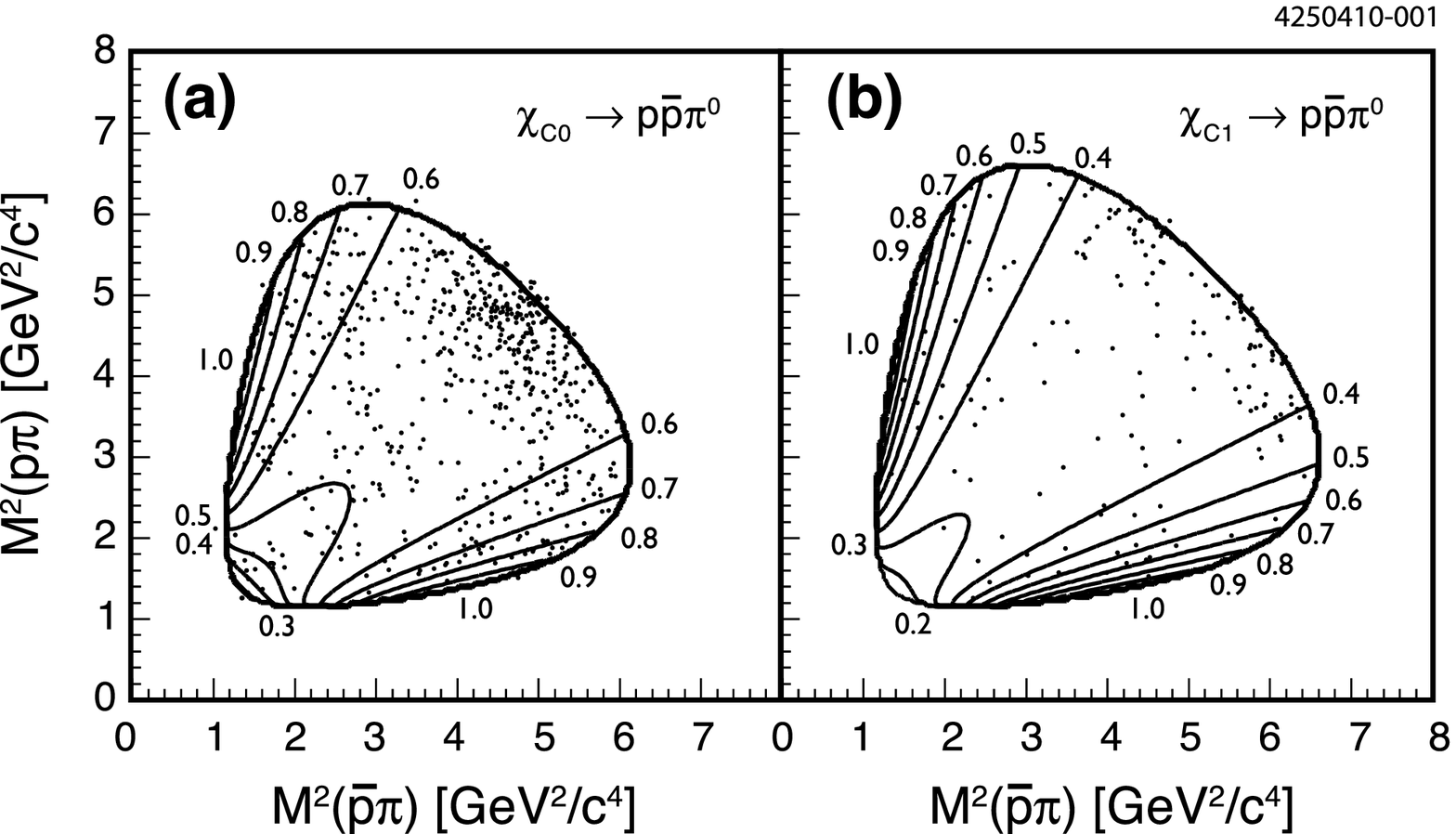,width=10.0cm}}
\vspace*{8pt}
\caption{Dalitz plots for (a)~$\chi_{c0}\to p\overline{p}\pi^0$ and (b)~$\chi_{c1}\to p\overline{p}\pi^0$.  The contours show the predictions from the meson emission model.\label{fig:pppi}}
\end{figure}

Starting with the full sample of 25.9~million $\psi(2S)$ decays, CLEO-c measured the branching fractions for $\chi_{cJ}\to p\overline{p}\pi^0$, $p\overline{p}\eta$, and $p\overline{p}\omega$~\cite{iupp}.  These results are important inputs to predictions of the related charmonium production rates ($p\overline{p}\to(c\overline{c})+X$, where $X$ is a hadron) that will be studied by the PANDA experiment at GSI.  In addition, the Dalitz plots of $\chi_{c0,1}\to p\overline{p}\pi^0$ have been studied (Figure~\ref{fig:pppi}).  An enhancement in $p\overline{p}$ mass appears, which is inconsistent with both predictions from the meson emission model~\cite{emission} and the $X(1870)$ threshold enhancement~\cite{besx}. 

\section{Study of $\psi(2S)$ decays to $\gamma p\overline{p}$, $p\overline{p}\pi^0$, and $p\overline{p}\eta$ and Search for $p\overline{p}$ Threshold Enhancements}

In addition to the $\chi_{cJ}$ decays discussed above, CLEO-c has also measured the branching fractions of $\psi(2S)$ decays to final states involving a $p\overline{p}$ pair~\cite{nupp}.  The Dalitz plots of $p\overline{p}\pi^0$ and $p\overline{p}\eta$ show evidence for the production of a number of $N^*$ resonances in $\pi^0p$ and $\eta p$, but do not show evidence for a $p\overline{p}$ threshold enhancement consistent with the $X(1870)$.  The $X(1870)$ is also not seen in the $\psi(2S)$ radiative decay to $p\overline{p}$.  However, evidence is found for the $X(1870)$ through the process $\psi(2S)\to\pi^+\pi^-J/\psi; J/\psi\to\gamma X(1870); X(1870)\to p\overline{p}$, which provides confirmation of the original BESII discovery~\cite{besx}.

\section{Search for $\psi(2S)\to\gamma\eta_c(2S)$ via Fully Reconstructed $\eta_c(2S)$ Decays}

In the spectroscopy of charmonium below $D\overline{D}$ threshold, the $\eta_c(2S)$ remains one of the least measured states.  CLEO-c performed a search for the $\eta_c(2S)$ in the allowed $M1$ transition, $\psi(2S)\to\gamma\eta_c(2S)$, where the potential $\eta_c(2S)$ was fully reconstructed in sixteen of its possible decay modes to light quark systems~\cite{etac}.  Upper limits on the product branching fractions, $B_1(\psi(2S)\to\gamma\eta_c(2S))\times B_2(\eta_c(2S)\to X)$, were separately set for each possible decay mode of the $\eta_c(2S)$, $X$.  The upper limits ranged from approximately $4\times10^{-6}$ to $4\times10^{-5}$.

\section{Higher-order Multipole Amplitudes in Charmonium Radiative Transitions}

\begin{figure}[pb]
\label{fig:multipoles}
\centerline{\psfig{file=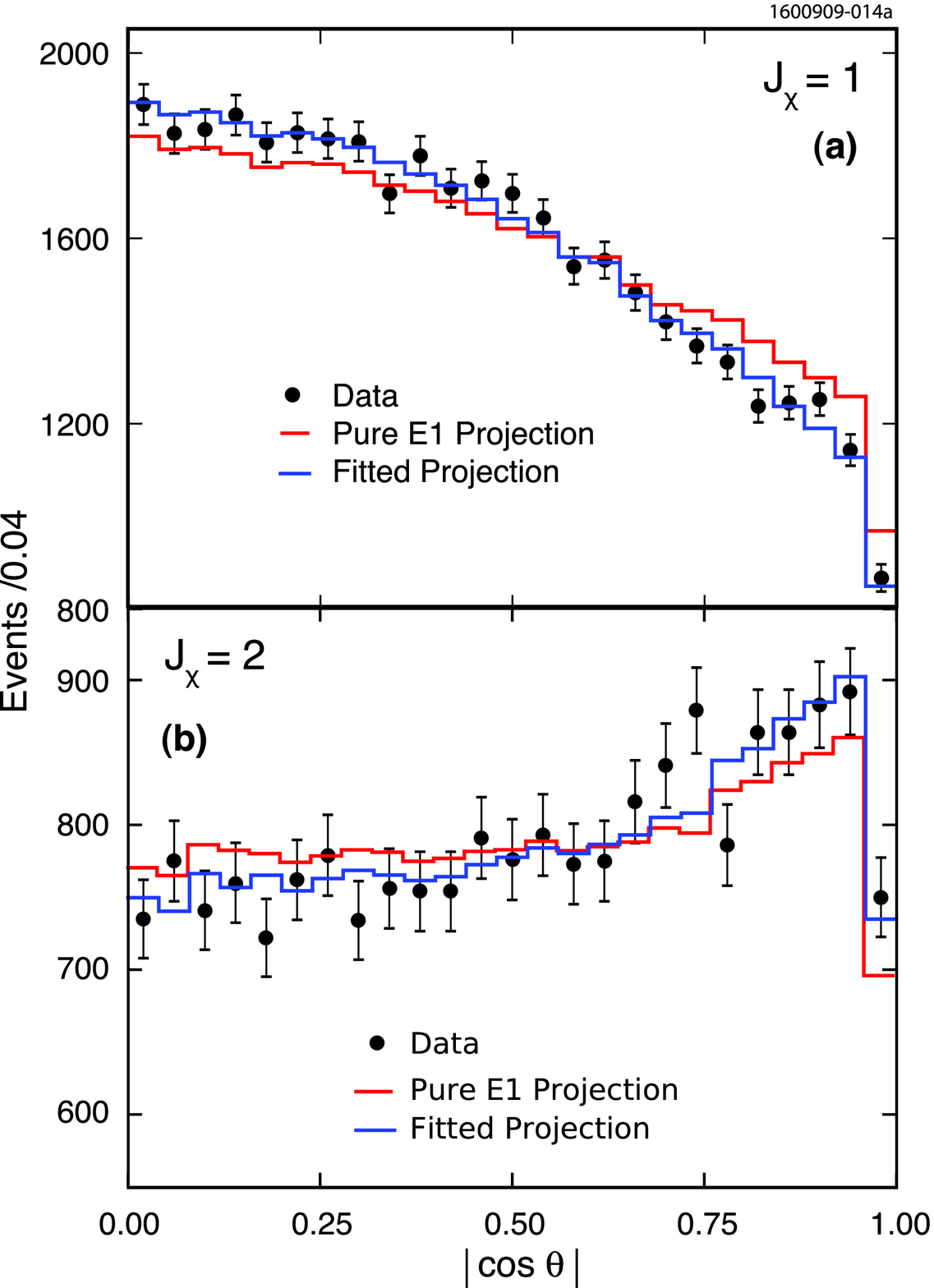,width=6.0cm}\psfig{file=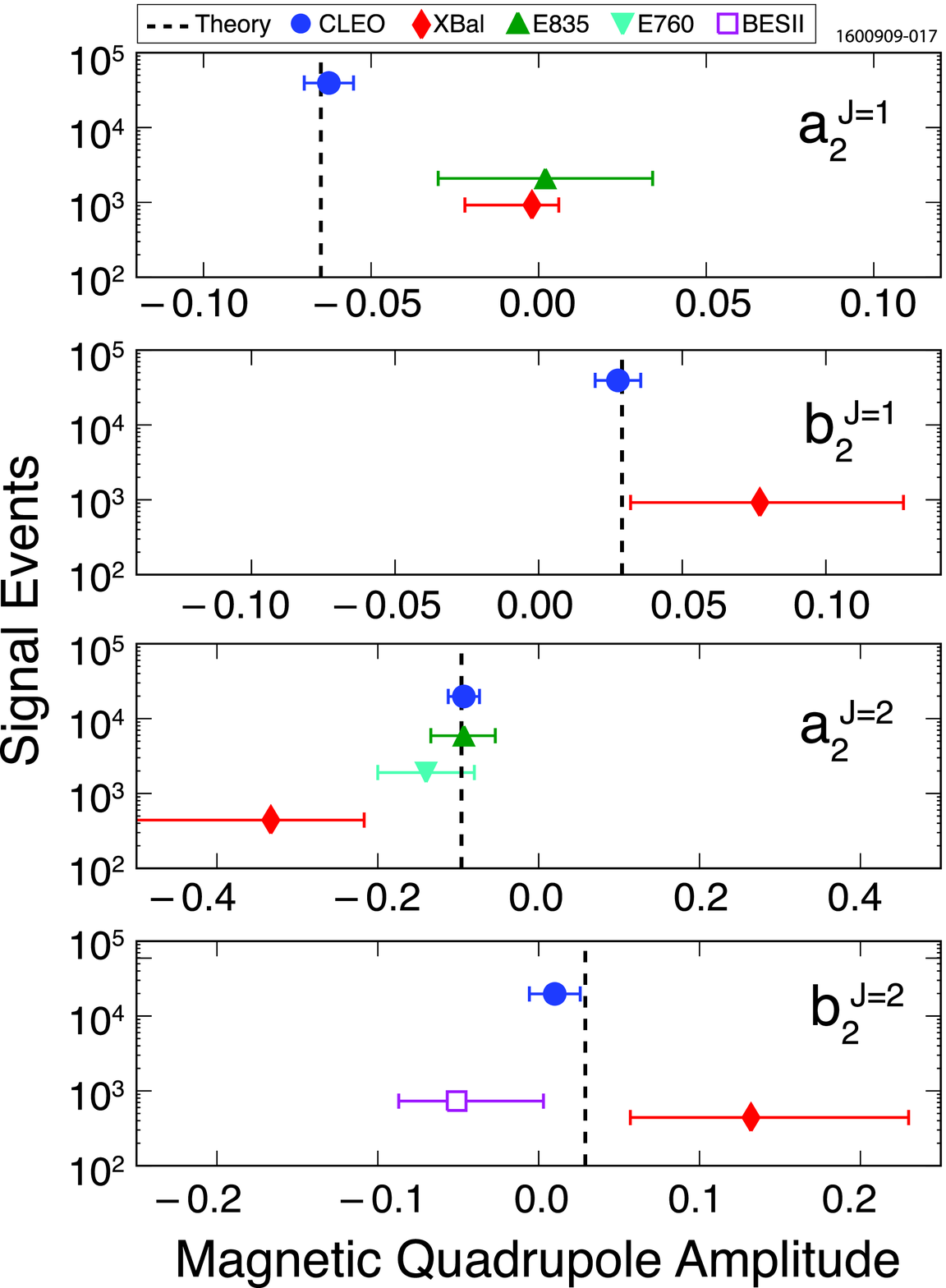,width=6.0cm}}
\vspace*{8pt}
\caption{Higher-Order Multipoles in $\psi(2S)\to\gamma\chi_{c1,2}$ and $\chi_{c1,2}\to\gamma J/\psi$.  (left) The cosine of the angle between the $l^+$ and the $\gamma$ from the $\chi_{cJ}$ decay in the $J/\psi$ rest frame from the process $\psi(2S)\to\gamma\chi_{cJ};\chi_{cJ}\to\gamma J/\psi;J/\psi\to l^+l^- (l=e,\mu)$ for (top) $J=1$ and (bottom) $J=2$.  This angle is sensitive to the M2 multipole.  The darker (blue) lines show improved fits when M2 amplitudes are included.  (right) The CLEO results (circles) for the four normalized M2 amplitudes compared to previous measurements (other points) and theoretical calculations assuming a zero anomalous magnetic moment of the charm quark and a charm quark mass of $1.5~\gevcc$ (dashed lines).}
\end{figure}

The radiative decays $\psi(2S)\to\gamma\chi_{c1,2}$ and
$\chi_{c1,2}\to\gamma J/\psi$ are dominated by electric dipole~(E1)
amplitudes, but are expected to have small additional
contributions from the higher-order magnetic quadrupole~(M2)
amplitudes. The relative contributions of the M2 amplitudes are, at least in principle, sensitive to the anomalous magnetic moment of the charm quark.  CLEO performed a high-statistics analysis of this process using the decay chains $\psi(2S)\to\gamma\chi_{cJ};\chi_{cJ}\to\gamma J/\psi;
J/\psi\to l^+l^- (l=e,\mu)$ for $J=1,2$, using clean samples of approximately
$40\,000$ events for $J=1$ and approximately $20\,000$ events for
$J=2$~\cite{multipoles}, significantly larger event samples than previous measurements.  Using an unbinned maximum likelihood fit to angular distributions,
CLEO found the following normalized M2 admixtures: for
$\psi(2S)\to\gamma\chi_{c1}$ $(\equiv b^{J=1}_2)$,
$(2.76\pm0.73\pm0.23)\times10^{-2}$; for $\psi(2S)\to\gamma\chi_{c2}$
$(\equiv b^{J=2}_2)$, $(1.0\pm1.3\pm0.3)\times10^{-2}$; for
$\chi_{c1}\to\gamma J/\psi$ $(\equiv a^{J=1}_2)$,
$(-6.26\pm0.63\pm0.24)\times10^{-2}$; and for $\chi_{c2}\to\gamma
J/\psi$ $(\equiv a^{J=2}_2)$, $(-9.3\pm1.6\pm0.3)\times10^{-2}$. For
the quoted $J=2$ measurements, the electric octupole~(E3) moments were
fixed to zero.  There is clear evidence for non-zero M2 amplitudes (left plot of Figure~\ref{fig:multipoles}).  These new measurements agree well with theoretical
expectations when the anomalous magnetic moment of the charm quark is
assumed to be zero and the mass of the charm quark is assumed to be
$1.5~\gevcc$, as shown in the right plot of Figure~\ref{fig:multipoles}.

\section{Transitions to the $h_c$ Above $D\overline{D}$ Threshold}

Finally, CLEO-c has presented preliminary results on the observation of $e^+e^-\to\pi^+\pi^-h_c$ at a center of mass energy of $4170~\mev$.  This builds upon a previous analysis of CLEO-c to study the $e^+e^-\to XJ/\psi$ $(X=\pi^+\pi^-, \pi^0\pi^0, K^+K^-, etc.)$ cross sections as a function of center of mass energy~\cite{xjpsi}.  In that analysis a rise in the cross section at $4260~\mev$ for both the $\pi^+\pi^-J/\psi$ and $\pi^0\pi^0J/\psi$ reactions was attributed to the $Y(4260)$, a charmonium hybrid meson candidate first discovered by BaBar~\cite{babary}.  Since that analysis, CLEO-c has collected a total of $586~\ipb$ of data at $4170~\mev$, about twenty times larger than the data sets used to search for $XJ/\psi$ between 4120 and 4200~$\mev$.  Using this larger data set, CLEO-c has made the first observation of $e^+e^-\to\pi^+\pi^-h_c$ at $4170~\mev$ and has searched for $e^+e^-\to\pi^+\pi^-h_c$ in the original scan data, which spans energies from $3970$ to $4260~\mev$.

\begin{figure}[pb]
\centerline{\psfig{file=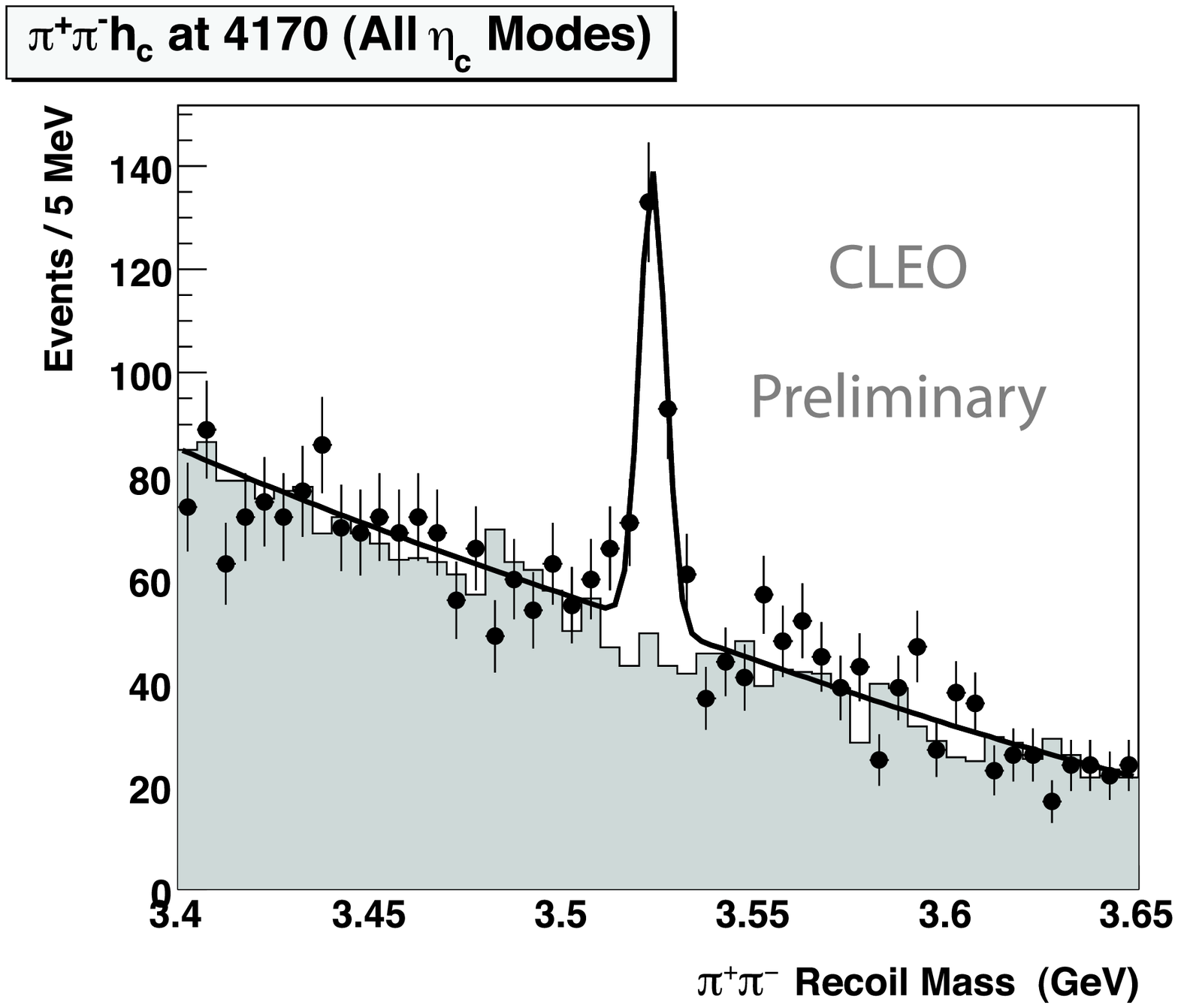,width=6.5cm}\psfig{file=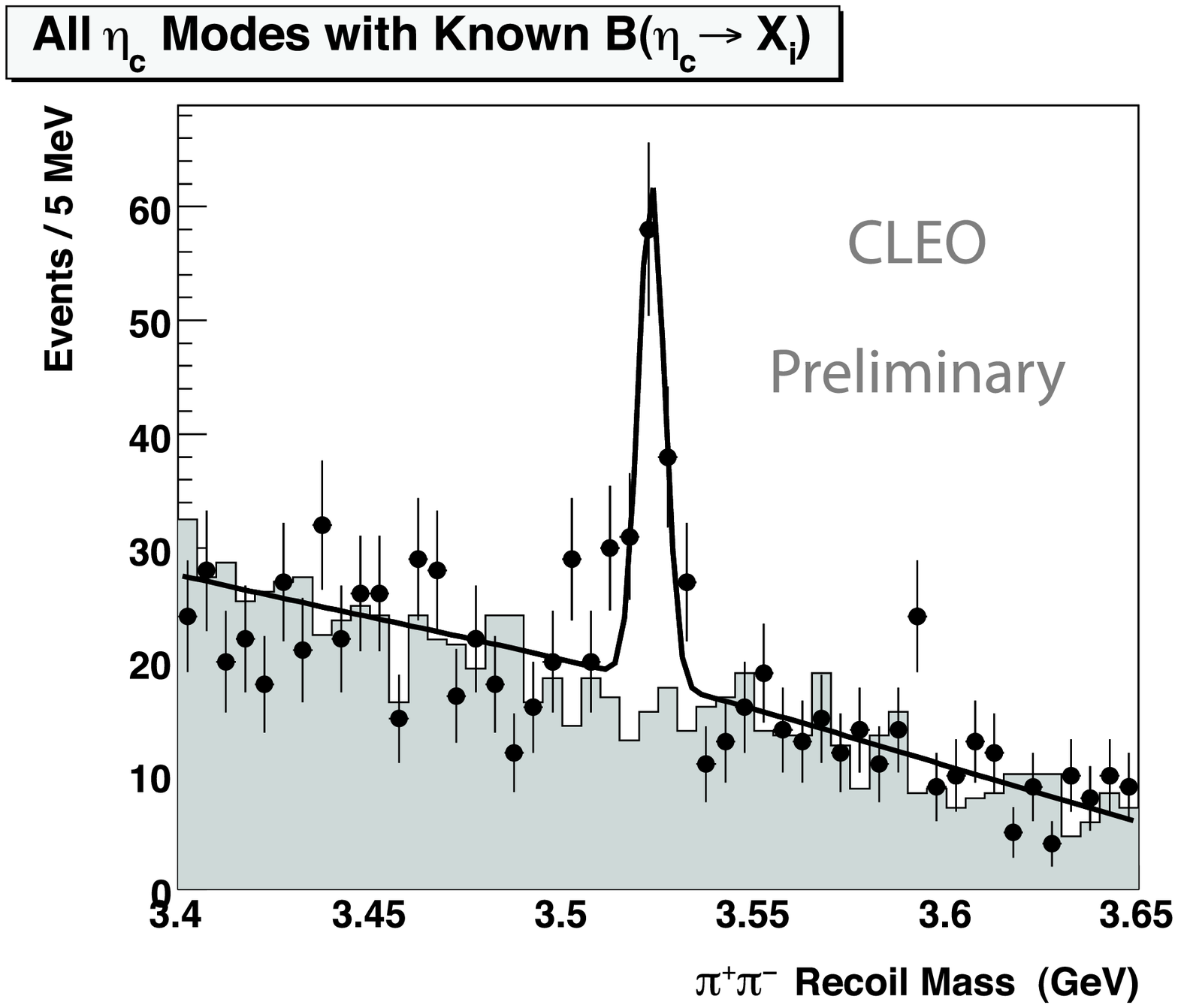,width=6.5cm}}
\vspace*{8pt}
\caption{The $\pi^+\pi^-$ recoil mass showing evidence for the reaction $e^+e^-\to\pi^+\pi^-h_c$ at a center of mass energy of $4170~\mev$ (preliminary). The points are data that satisfied the kinematic fit (with a good $\chi^2$); the shaded regions are data that had a poorer kinematic fit (in the $\chi^2$ sideband region).  (left)~The sum of all twelve exclusive decay modes of the $\eta_c$; (right)~the subset of those twelve modes that have known branching fractions.\label{fig:hcsignal}}
\end{figure}

The reaction $e^+e^-\to\pi^+\pi^-h_c$ was exclusively reconstructed using $h_c\to\gamma\eta_c$ and twelve decay modes of the $\eta_c$ ($\pi^+\pi^+\pi^-\pi^-$, $\pi^+\pi^+\pi^-\pi^-\pi^0\pi^0$, $\pi^+\pi^+\pi^+\pi^-\pi^-\pi^-$, $K^{\pm}K_S\pi^{\mp}$, $K^{\pm}K_S\pi^{\mp}\pi^+\pi^-$, $K^+K^-\pi^0$, $K^+K^-\pi^+\pi^-$, $\eta\pi^+\pi^-$, $\eta\pi^+\pi^+\pi^-\pi^-$, $K^+K^-\pi^+\pi^-\pi^0$, $K^+K^-\pi^+\pi^+\pi^-\pi^-$, $K^+K^+K^-K^-$).  All final state particles were detected and the sum of their four-momenta was constrained to the four-momentum of the initial $e^+e^-$ system.  The quality ($\chi^2$) of this kinematic fit was used to select events.  The $\eta_c$ was selected by requiring the recoil mass of the $\gamma\pi^+\pi^-$ system be within $50~\mevcc$ of the $\eta_c$ mass.  Then the $\pi^+\pi^-$ recoil mass was used to isolate the $\pi^+\pi^-h_c$ signal.

The left plot of Figure~\ref{fig:hcsignal} shows the $\pi^+\pi^-$ recoil mass when all twelve decay modes of the $\eta_c$ are summed together.  There is a peak at the $h_c$ mass, which contains $150\pm17$ events and has a $9.4~\sss$ significance.  To convert this to a cross section, however, one must know the $\eta_c$ branching fractions.  The right plot of Figure~\ref{fig:hcsignal} shows the same figure except now restricted to those decay modes of the $\eta_c$ that have measured branching fractions.  Here there are $74\pm11$ events with a $7.7~\sss$ significance.  This corresponds to a cross section of $e^+e^-\to\pi^+\pi^-h_c$ of $12.2\pm1.8\pm2.4\pm4.0~\pb$, where the first error is statistical, the second is systematic, and the third is due to the errors on the external measurements of the $\eta_c$ branching fractions and the branching fraction of $h_c\to\gamma\eta_c$.  For reference, the cross section for $e^+e^-\to\pi^+\pi^-J/\psi$ at $4170~\mev$ is $8\pm2\pm2~\pb$.

\begin{figure}[pb]
\centerline{\psfig{file=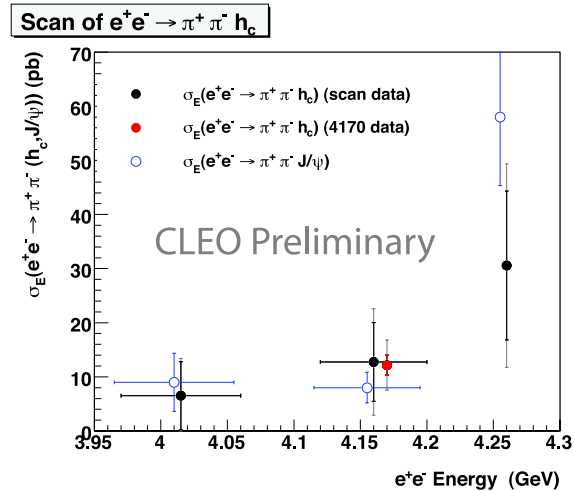,width=8.2cm}}
\vspace*{8pt}
\caption{Scan of the $\pi^+\pi^-h_c$ cross section as a function of center of mass energy (preliminary). \label{fig:scan}}
\end{figure}

To obtain a scan of the $\pi^+\pi^-h_c$ cross section as a function of center of mass energy, the same procedure was followed using the smaller samples of scan data.  Following the $XJ/\psi$ scan~\cite{xjpsi}, the data is divided into three bins: $3970 - 4060$; $4120 - 4200$; and $4260~\mev$.  Figure~\ref{fig:scan} shows the resulting $\pi^+\pi^-h_c$ cross section as a function of center of mass energy. The $\pi^+\pi^-J/\psi$ results are included for reference.  There is perhaps a suggestive rise at $4260~\mev$, but more data is clearly needed.

\end{document}